\documentclass[aps,twocolumn,showpacs]{revtex4-1}
\usepackage{amsmath}
\usepackage{epsfig}

\begin{document}

\title{Full heavy dibaryons}

\author{Hongxia Huang$^1$}\email[E-mail: ]{hxhuang@njnu.edu.cn}
\author{Jialun Ping$^1$}\email[E-mail: ]{jlping@njnu.edu.cn (Corresponding author)}
\author{Xinmei Zhu$^2$}\email[E-mail: ]{zxm$\_$yz@126.com}
\author{Fan Wang$^3$}\email[E-mail: ]{fgwang@foxmail.com}
\affiliation{$^1$Department of Physics, Nanjing Normal University, Nanjing, Jiangsu 210097, China}
\affiliation{$^2$Department of Physics, Yangzhou University, Yangzhou 225009, P. R. China}
\affiliation{$^3$Department of Physics, Nanjing University,
Nanjing 210093, P.R. China}

\begin{abstract}
The existence of full heavy dibaryons $\Omega_{ccc}\Omega_{bbb}$, $\Omega_{ccc}\Omega_{ccc}$ and $\Omega_{bbb}\Omega_{bbb}$ with $J=0,~1,~2,~3$ and $P=\pm1$ are investigated in the framework of a constituent quark model. The dibaryon composed of six $c$ or $b$ quarks with $J^{P}=0^{+}$ is possible to be bound, while the dibaryon with the $cccbbb$ configuration is difficult to form any bound state. We also study the interaction between two full heavy baryons, and the effective potentials, suggesting the existence of dibaryons $\Omega_{ccc}\Omega_{ccc}$ and $\Omega_{bbb}\Omega_{bbb}$, and the absence of binding for $\Omega_{ccc}\Omega_{bbb}$ system. Besides, the calculation of the low-energy scattering phase shifts of the $\Omega_{ccc}\Omega_{ccc}$ and $\Omega_{bbb}\Omega_{bbb}$, as well as the scattering length also confirm the existence of stable full heavy dibaryons $\Omega_{ccc}\Omega_{ccc}$ and $\Omega_{bbb}\Omega_{bbb}$.
\end{abstract}


\maketitle

\setcounter{totalnumber}{5}

\emph{Introduction.}--- Understanding the hadron-hadron interactions and searching for exotic quark states are important topics in temporary hadron physics, among which questing for dibaryons is a long-standing challenge.
Among the various theoretical study of dibaryons, the lattice QCD simulations by HAL QCD Collaboration have studied the six-quark systems containing light or strange quarks and confirmed the existence of the $N\Omega$ and $\Omega\Omega$ bound states with nearly physical quark masses ($m_{\pi}\simeq 146$ MeV and $m_{K}\simeq 525$ MeV)~\cite{Lattice1,Lattice2}. Recently, Junnarkar and Mathur reported the first lattice QCD study of deuteron-like ($np$-like) dibaryons with heavy quark flavors by using a state-of-the-art lattice QCD calculation~\cite{Lattice3}. They suggested that the dibaryons $\Omega_{c}\Omega_{cc} (sscscc)$, $\Omega_{b}\Omega_{bb} (ssbsbb)$, and $\Omega_{ccb}\Omega_{cbb} (ccbcbb)$ were stable under strong and electromagnetic interactions, and they also found that the binding of these dibaryons became stronger as they became heavier in mass. However, the distinct conclusion was claimed in the work of Ref.~\cite{Richard}, where the authors explored the possibility of very heavy dibaryons with three charm quarks and three beauty quarks ($bbbccc$) in potential models, and concluded that there was no evidence for any stable state in such very heavy flavored six-quark system. Moreover, in the case of equal masses $m_{b} = m_{c}$, there was no bound state either in their calculation~\cite{Richard}. Thus, the stability of the very heavy dibaryons has been largely undecided. It may due to a lack of experimental information about the strength of the interaction between heavy quarks.

The discovery of the doubly charmed baryon $\Xi_{cc}$ by the LHCb Collaboration~\cite{LHCb1} provided the crucial experimental input for the existence of the stable heavy tetraquark $QQ\bar{q}\bar{q}$~\cite{Karliner}. E. J. Eichten and C. Quigg also predicted the existence of the novel narrow doubly heavy tetraquark $QQ\bar{q}\bar{q}$ by using the relations derived from the heavy-quark symmetry~\cite{Eichten}. Inspired by the LHCb Collaboration's observation of the hidden-charm $P_{c}$ pentaquarks~\cite{LHCb2}, Ref.~\cite{LiuM} connected the properties of heavy antimesons ($D$, $D^{*}$) and doubly heavy baryons ($\Xi_{cc}$, $\Xi^{*}_{cc}$) with the application of heavy-antiquark-diquark symmetry (HADS)~\cite{Savage}, which states that the two heavy quarks within a doubly heavy baryon behave approximately as a heavy antiquark, and predicted the dibaryons $0^{+}$ $\Xi_{cc}\Sigma_{c}$, $1^{+}$ $\Xi_{cc}\Sigma^{*}_{c}$, $2^{+}$ $\Xi^{*}_{cc}\Sigma_{c}$, and $3^{+}$ $\Xi^{*}_{cc}\Sigma^{*}_{c}$. Very recently, the LHCb collaboration reported their results on the observations of full-charm states ($cc\bar{c}\bar{c}$). A narrow structure $X(6900)$, matching the lineshape of a resonance and a broad structure next to the $di-J/\psi$ mass threshold was obtained~\cite{LHCb3}. Such a breakthrough offers more information for the searching of the tetraquark consisting of four charm quarks. By applying the HADS, a full heavy dibaryon $cccccc$ is also possible to be a stable state.

Thus, the observation of the full heavy dibaryons become more interesting. Based on the study of strange dibaryon $N\Omega$, we predicted the $N\Omega$-like dibaryons $N\Omega_{ccc}$ and $N\Omega_{bbb}$ in the framework of the constituent quark model~\cite{Huang1}. The study of the $H$-particle was extended to the heavy sector states $\Lambda_{c}\Lambda_{c}$ and $\Lambda_{b}\Lambda_{b}$~\cite{Huang2}. In the present Letter, we extend our study to the possibility of the full heavy dibaryon systems. We investigate the interaction between two full heavy baryons. The effective potential, binding energy, as well as the low-energy scattering phase shifts, the scattering length, and the effective range are also calculated to explore the existence of the full heavy dibaryon systems.

\emph{Model and method.}--- We study the full heavy dibaryon systems within the constituent quark model, which is similar to what was used in our previous work of dibaryons: $d^{*}$, $N\Omega$, and $\Omega\Omega$~\cite{Ping1,Huang3,Huang4}; and it has been extended to the heavy dibaryons $N\Omega_{ccc}$~\cite{Huang1}. When applying to the full heavy systems, neither the $SU(3)$ scalar octet meson exchange nor the Goldstone-boson exchange works. Besides, since we investigate the ground state of the full heavy systems as the first step, the spin-orbit and tensor forces are neglected. Here, we list the interaction Hamiltonian we used in this work:
\begin{equation}
H=\sum_{i=1}^6\left(m_i+\frac{p_i^2}{2m_i}\right)-T_{\rm{CM}} +\sum_{j>i=1}^6
\left(V^{\rm{CON}}_{ij}+V^{\rm{OGE}}_{ij}\right),
\end{equation}
Where $T_{\rm{CM}}$ is the kinetic energy of the center of mass. $V^{\rm{CON}}_{ij}$ is the phenomenological confinement potential:
\begin{equation}
V^{\rm{CON}}_{ij}=-a_{c} \boldsymbol{\lambda}^c_{i}\cdot \boldsymbol{
\lambda}^c_{j} ({r^2_{ij}}+v_{0}),
\end{equation}
and $V^{\rm{OGE}}_{ij}$ is the one-gluon-exchange interaction:
\begin{equation}
V^{\rm{OGE}}_{ij}=\frac{\alpha_s }{4}\boldsymbol{\lambda}^{c}_i \cdot
\boldsymbol{\lambda}^{c}_j
\left[\frac{1}{r_{ij}}-\frac{\pi}{2}\delta(\boldsymbol{r}_{ij})(\frac{1}{m^2_i}+\frac{1}{m^2_j}
+\frac{4\boldsymbol{\sigma}_i\cdot\boldsymbol{\sigma}_j}{3m_im_j})\right].
\end{equation}
The other symbols in the above expressions have their usual meanings. All parameters, which are fixed by fitting to the masses of baryons 
with light flavors and heavy flavors, are taken from our previous work~\cite{Huang1}.

The resonating group method (RGM)~\cite{RGM} and generating coordinates method~\cite{GCM} are used to carry out a dynamical calculation. 
The conventional ansatz for the two-cluster wavefunctions is
\begin{equation}
\psi_{6q} = {\cal A }\left[[\phi_{B_{1}}\phi_{B_{2}}]^{[\sigma]IS}\otimes\chi_{L}(\boldsymbol{R})\right]^{J}, \label{6q}
\end{equation}
where the symbol ${\cal A }$ is the anti-symmetrization operator. For the dibaryon with six $c$ quarks or six $b$ quarks, ${\cal A } = 1-9P_{36}$; 
while for the dibaryon composed of three $c$ quarks and three $b$ quarks, ${\cal A } = 1$. $[\sigma]=[222]$ gives the total color symmetry and all 
other symbols have their usual meanings. $\phi_{B_{i}}$ is the 3-quark cluster wavefunction, and $\chi_{L}(\boldsymbol{R})$ is the relative motion 
wavefunction, which is expanded by a set of gaussians with different centers,
\begin{eqnarray}
& & \chi_{L}(\boldsymbol{R}) = \frac{1}{\sqrt{4\pi}}(\frac{3}{2\pi b^2})^{3/4} \sum_{i=1}^{n} C_{i}  \nonumber \\
&& ~~~~\times  \int \exp\left[-\frac{3}{4b^2}(\boldsymbol{R}-\boldsymbol{S}_{i})^{2}\right] Y_{LM}(\hat{\boldsymbol{S}_{i}})d\hat{\boldsymbol{S}_{i}},~~~~~
\end{eqnarray}
By including the center of mass motion:
\begin{equation}
\phi_{C} (\boldsymbol{R}_{C}) = (\frac{6}{\pi b^{2}})^{3/4}e^{-\frac{3\boldsymbol{R}^{2}_{C}}{b^{2}}},
\end{equation}
the ansatz Eq.(\ref{6q}) can be rewritten as
\begin{eqnarray}
& &\psi_{6q} = {\cal A} \sum_{i=1}^{n} C_{i} \int \frac{d\hat{\boldsymbol{S}_{i}}}{\sqrt{4\pi}}
\prod_{\alpha=1}^{3}\phi_{\alpha}(\boldsymbol{S}_{i}) \prod_{\beta=4}^{6}\phi_{\beta}(-\boldsymbol{S}_{i}) \nonumber \\
& & ~~~~\times   \left[[\eta_{I_{1}S_{1}}(B_{1})\eta_{I_{2}S_{2}}(B_{2})]^{IS}Y_{LM}(\hat{\boldsymbol{S}_{i}})\right]^{J} \nonumber \\
& & ~~~~\times  [\chi_{c}(B_{1})\chi_{c}(B_{2})]^{[\sigma]}, \label{6q2}
\end{eqnarray}
where $\phi_{\alpha}(\boldsymbol{S}_{i})$ and $\phi_{\beta}(-\boldsymbol{S}_{i})$ are the single-particle orbital wavefunctions with different 
reference centers:
\begin{eqnarray}
\phi_\alpha(\boldsymbol {S_{i}})=\left(\frac{1}{\pi
b^2}\right)^{\frac{3}{4}}e^ {-\frac{(\boldsymbol {r}_{\alpha}-\boldsymbol
{S_i}/2)^2}{2b^2}},
 \nonumber\\
\phi_\beta(-\boldsymbol {S_{i}})=\left(\frac{1}{\pi
b^2}\right)^{\frac{3}{4}}e^ {-\frac{(\boldsymbol {r}_{\beta}+\boldsymbol
{S_i}/2)^2}{2b^2}} .
\end{eqnarray}
From the variational principle, one generalized eigenvalue equation:
\begin{equation}
\sum_{j} C_{j}H_{i,j}= E \sum_{j} C_{j}N_{i,j}.
\end{equation}
where $H_{i,j}$ and $N_{i,j}$ are the Hamiltonian matrix elements and overlaps, respectively. By solving the generalized eigen problem, 
we can obtain the energy and the corresponding wavefunctions of the dibaryon system.

\emph{Results.}--- The masses of the full heavy baryon $\Omega_{ccc}$ and $\Omega_{bbb}$ have been calculated in our former work~\cite{Huang1}, 
and they are $5068.8$ MeV and $15111.6$ MeV, respectively. So the threshold of $\Omega_{ccc}\Omega_{ccc}$, $\Omega_{bbb}\Omega_{bbb}$, and 
$\Omega_{ccc}\Omega_{bbb}$ is $10137.6$ MeV, $30223.2$ MeV,and $20180.4$ MeV, respectively. We first calculate the energy of these three full 
heavy dibaryon systems with the quantum numbers $J=0,~1,~2$, and $3$. Since both the $\Omega_{ccc}\Omega_{ccc}$ and $\Omega_{bbb}\Omega_{bbb}$ 
include six identical quarks, the $J=0$ and $2$ systems have the positive parity while $J=1$ and $3$ systems have the negative parity. 
However, nonidentical quarks in two subclusters of the system $\Omega_{ccc}\Omega_{bbb}$ leads to both positive and negative parity are possible 
for this system with $J=0,~1,~2$, and $3$. The details are given in Table~\ref{dibaryons}.
\begin{table}[ht]
\caption{The energy (in MeV) of the full heavy dibaryon systems.}
\begin{tabular}{lccccccccc}
\hline\hline
 ~~~~~~$J^{P}$~~~~~~ & ~~~~$0^{+}$~~~~ & ~~~~$1^{-}$~~~~ & ~~~~$2^{+}$~~~~  &  ~~~~$3^{-}$~~~~   \\ \hline
 ~~$\Omega_{ccc}\Omega_{ccc}$ & 10136.3 & 10146.5 & 10139.5  & 10147.8  \\
 ~~$\Omega_{bbb}\Omega_{bbb}$ & 30222.9 & 30226.2 & 30223.8  & 30226.6  \\ \hline
 ~~~~~~$J^{P}$~~~~~~            & ~~~$0^{\pm}$~~~ & ~~~$1^{\pm}$~~~ & ~~~$2^{\pm}$~~~  &  ~~~$3^{\pm}$~~~   \\ \hline
~~$\Omega_{ccc}\Omega_{bbb}$ & 20181.4 & 20181.4 & 20181.4  & 20181.4  \\
\hline\hline
\end{tabular}
\label{dibaryons}
\end{table}

Obviously, the energies of all states are above the corresponding theoretical threshold, except the state $\Omega_{ccc}\Omega_{ccc}$ and 
$\Omega_{bbb}\Omega_{bbb}$ with $J^{P}=0^{+}$, the energy of which is $-1.3$ MeV and $-0.3$ MeV lower than the corresponding thresholds, 
respectively. Therefore, for the $J^{P}=0^{+}$ systems, the dibaryon with six $c$ quarks or six $b$ quarks is possible to form bound state, 
while the one composed of three $c$ quarks and three $b$ quarks is unbound. This is due to the different symmetry requirement for the 
different dibaryons. It is antisymmetric between the same baryon clusters, while there is no symmetry requirement between the clusters 
$\Omega_{ccc}$ and $\Omega_{bbb}$. We also note that the energy of the states $\Omega_{ccc}\Omega_{bbb}$ with different quantum numbers is 
the same. The main reason is that there is no interaction between two color singlet subclusters due to no exchange term.

To investigate the interactions between two full heavy baryons in detail, we derive the analytical expressions of the Hamiltonian matrix 
elements of each interaction and list them in Table~\ref{inter}. To save space, we only show the results of the systems with $J^{P}=0^{+}$ here. 
Since the expressions of the dibaryons with six $c$ quarks and six $b$ quarks are the same, we only list the expressions of two systems, which are $\Omega_{ccc}\Omega_{ccc}$ and $\Omega_{ccc}\Omega_{bbb}$. All the results shown in the table are ones with subtracting the internal interaction 
of the two baryons. Note that when two clusters are far apart ($S\rightarrow \infty$), the kinetic energy of the relative motion between two 
$\Omega_{ccc}$ clusters is $3(4m_{c}b^{2})^{-1}$, and the one between $\Omega_{ccc}$ and $\Omega_{bbb}$ is 
$\frac{3}{8b^{2}}(m_{c}^{-1}+m_{b}^{-1})$.
From the analytical expressions in Table~\ref{inter} we can see that there is no confinement interaction between two subclusters. 
For $\Omega_{ccc}\Omega_{bbb}$ system, the color Coulomb interaction or the color magnetic interaction between the baryons $\Omega_{ccc}$ 
and $\Omega_{bbb}$ is also 0. The kinetic energy between $\Omega_{ccc}$ and $\Omega_{bbb}$ is only the one of the relative motion, 
which indicates that there is no attractive interaction between $\Omega_{ccc}$ and $\Omega_{bbb}$, so it is difficult for the 
$\Omega_{ccc}\Omega_{bbb}$ system to form any bound state. By contrast, the case is different for the $\Omega_{ccc}\Omega_{ccc}$ system. 
The kinetic energy will be negative by subtracting the one of the relative motion at $S\rightarrow \infty$, which means that it can provide 
effective attractive between two $\Omega_{ccc}$ subclusters. Besides, the Coulomb interaction will be negative with some suitable value of $S$, 
although the color magnetic interaction is positive. So it is possible for the $\Omega_{ccc}\Omega_{ccc}$ system to form a bound state.
\begin{table}
\caption{The matrix elements (in MeV) of each interaction term in Hamiltonian for the full heavy dibaryon systems $\Omega_{ccc}\Omega_{ccc}$ 
and $\Omega_{ccc}\Omega_{bbb}$ with $J^{P}=0^{+}$. $V_{k}$: kinetic energy; $V_{\rm{CON}}$: confinement; $V_{CL}$: color Coulomb and
$V_{\rm{CMI}}$: color magnetic interaction in the OGE. $O=\exp[-(\frac{S^{2}}{2b^{2}})]$, $x=S/\sqrt{8b^2}$; and $S$ is the distance between two baryons.}
\begin{tabular}{lccccccccc}
\hline\hline
 Terms & ~~~$\Omega_{ccc}\Omega_{ccc}$~~~ & ~~~$\Omega_{ccc}\Omega_{bbb}$~~~    \\ \hline
 $V_{k}$ & $\frac{1}{4m_{c}b^{2}}[3-\frac{S^{2}}{b^{2}}(\frac{O}{1+O})]$ & $\frac{3}{8b^{2}}(\frac{1}{m_{c}}+\frac{1}{m_{b}})$  \\
 $V_{\rm{CON}}$ & 0 &0  \\
 $V_{CL}$& $\alpha_{s}[\frac{4}{b}\sqrt{\frac{2}{\pi}}-\frac{32}{3S}\rm{erf}(x)+\frac{4}{3S}\rm{erf}(2x)](\frac{O}{1+O})$ & 0  \\
 $V_{\rm{CMI}}$& $(\frac{1}{2\pi b^{2}})^{3/2} \frac{\alpha_{s}\pi}{m^{2}_{c}}[\frac{4}{9}+\frac{80}{9}O^{1/4}+\frac{4}{3}O](\frac{O}{1+O})$ & 0 \\
\hline\hline
\end{tabular}
\label{inter}
\end{table}

\begin{figure}[ht]
\epsfxsize=3.0in \epsfbox{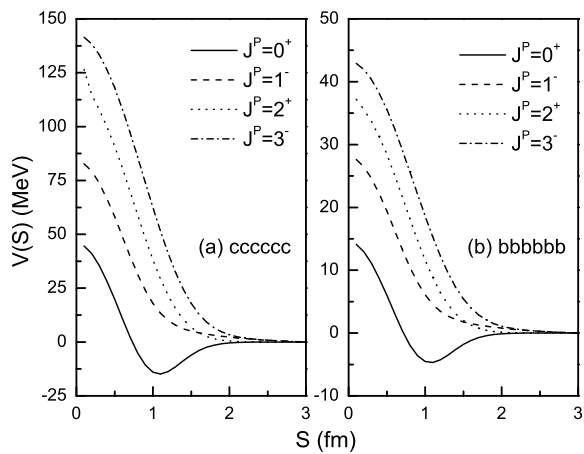} \vspace{-0.1in}

\caption{The effective potential of the $\Omega_{ccc}\Omega_{ccc}$ and $\Omega_{bbb}\Omega_{bbb}$ systems.}
\end{figure}

The effective potentials are also calculated to understand the interaction between two full heavy baryons, which are shown in 
Figs. 1 and 2.  The effective potential between two colorless clusters is defined as $V(S)=E(S)-E(\infty)$, where $E(S)$ is the diagonal matrix 
element of the Hamiltonian of the system in the generating coordinate. It is clear that the potential for both $\Omega_{ccc}\Omega_{ccc}$ and $\Omega_{bbb}\Omega_{bbb}$ with $J^{P}=0^{+}$ are attractive, which leads to the possibility for these two systems to form bound states. 
However, the potentials for other systems are all repulsive. That is why the other states are unbound.

\begin{figure}[ht]
\epsfxsize=3.0in \epsfbox{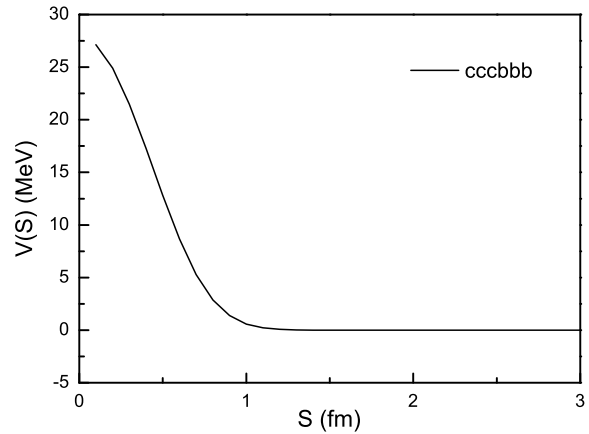} \vspace{-0.1in}

\caption{The effective potential of the $\Omega_{ccc}\Omega_{bbb}$ system.}
\end{figure}

To further check the possibility of the bound state composed of six $c$ quarks or six $b$ quarks, we also calculate the low-energy scattering 
phase shifts of the $S-$wave $\Omega_{ccc}\Omega_{ccc}$ and $\Omega_{bbb}\Omega_{bbb}$ systems, which are shown in Fig. 3. The well developed 
Kohn-Hulthen-Kato(KHK) variational method is used here, the details of which can be found in Ref.~\cite{RGM}. It is obvious that the scattering 
phase shifts of both two systems go to $180$ degrees at $E_{c.m.}\sim 0$ and rapidly decreases as $E_{c.m.}$ increases, suggesting the presence 
of a bound state.

\begin{figure}
\epsfxsize=3.0in \epsfbox{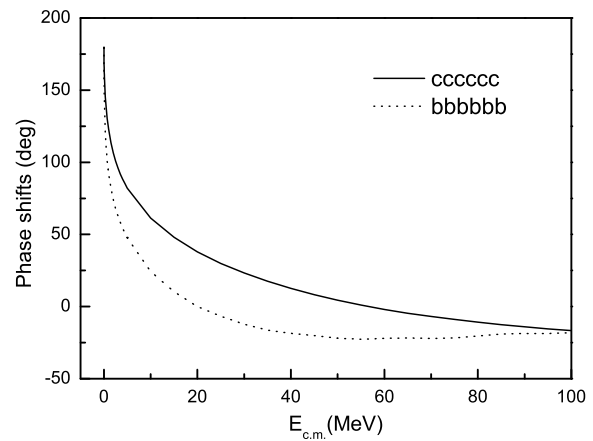} \vspace{-0.1in}

\caption{The phase shifts of the $S-$wave $\Omega_{ccc}\Omega_{ccc}$ and $\Omega_{bbb}\Omega_{bbb}$ systems with $J^{P}=0^{+}$.}
\end{figure}

Then, we can extract the low-energy scattering parameters, scattering length $a_{0}$ and the effective range $r_{0}$ of both 
$\Omega_{ccc}\Omega_{ccc}$ and $\Omega_{bbb}\Omega_{bbb}$ systems by using the formula:
\begin{eqnarray}
k\cot\delta & = & -\frac{1}{a_{0}}+\frac{1}{2}r_{0}k^{2}+{\cal O}(k^{4})
\end{eqnarray}
where $\delta$ is the low-energy phase shifts obtained above; $k$ is the momentum of relative motion with 
$k=\sqrt{2\mu E_{c.m.}}$, in which $\mu$ is the reduced mass of two baryons, and $E_{c.m.}$ is the incident energy.
Meanwhile, the binding energy $B^{\prime}$ can be calculated according to the relation:
\begin{eqnarray}
B^{\prime} & = &\frac{\hbar^2\alpha^2}{2\mu}
\end{eqnarray}
where $\alpha$ is the wave number which can be obtained from the relation~\cite{Babenko}:
\begin{eqnarray}
r_{0} & = &\frac{2}{\alpha}(1-\frac{1}{\alpha a_{0}})
\end{eqnarray}
Please note that we use another method to calculate the binding energy here, so we label it as $B^{\prime}$.
The results are listed in Table \ref{length}, from which we can see that the scattering length are all positive 
for both dibaryons $\Omega_{ccc}\Omega_{ccc}$ and $\Omega_{bbb}\Omega_{bbb}$, which implies the existence of a bound state of 
these two full-heavy dibaryons. The binding energies obtained here are coincident with those calculated from Table~\ref{dibaryons}.
\begin{center}
\begin{table}[h]
\caption{The scattering length $a_{0}$, effective range $r_{0}$,
and binding energy $B^{\prime}$ of the $\Omega_{ccc}\Omega_{ccc}$ and $\Omega_{bbb}\Omega_{bbb}$ systems with $J^{P}=0^{+}$.}
\begin{tabular}{lcccc}
\hline \hline
  & ~$a_{0}$~(fm)~ & ~$r_{0}$~(fm)~ & ~$B^{\prime}$~(MeV)~    \\ \hline
 ~~~$\Omega_{ccc}\Omega_{ccc}$~~~ & 3.1572  & 1.1729 & -1.3    \\ \hline
 ~~~$\Omega_{bbb}\Omega_{bbb}$~~~ & 3.7003 & 1.1644 & -0.3   \\ \hline
 \hline
\end{tabular}
\label{length}
\end{table}
\end{center}

\emph{Discussion and outlook.}--- The main conclusion of our dynamical investigation of full heavy dibaryons is that the dibaryons composed 
of six $c$ quarks or six $b$ quarks is possible to form bound states, and the quantum numbers are $J^{P}=0^{+}$. But there is no evidence for 
any stable dibaryon with the type of $\Omega_{ccc}\Omega_{bbb}$. This outcome is based on the traditional constituent quark model calculation. 
The principal reason for the instability of $\Omega_{ccc}\Omega_{bbb}$ system is that there is no symmetry requirement between the clusters 
$\Omega_{ccc}$ and $\Omega_{bbb}$ because the quark $c$ and quark $b$ are not identical quarks. While for the full heavy dibaryons 
$\Omega_{ccc}\Omega_{ccc}$ or $\Omega_{bbb}\Omega_{bbb}$, the requirement of the antisymmetrization between the same baryon clusters introduce 
attractive interaction between two full heavy baryons, which leads to a super-heavy bound dibaryon $\Omega_{ccc}\Omega_{ccc}$ or $\Omega_{bbb}\Omega_{bbb}$. 
This conclusion seems to be in tension with the results of Ref.~\cite{Richard}, which concluded that there was no evidence for any stable super-heavy
hexaquark of the $cccbbb$ and similar configurations, and the principal reason is the constraint of antisymmetrization in both the $c$ and the $b$ sectors. 
In Ref.~\cite{Richard}, a ``toy model" was proposed to calculate a configuration $cc^{'}c^{''}bb^{'}b^{''}$ with nonidentical $c-$type and $b-$type 
of quarks, and a binding energy of about $100$ MeV was obtained. We also study the same configuration $cc^{'}c^{''}bb^{'}b^{''}$ within the ``toy model", 
but no any stable is obtained. The main reason is still no symmetry requirement between the two clusters with nonidentical quarks.

To conclude, in the framework of the constituent quark model, the dibaryon composed of six $c$ quarks or six $b$ quarks is possible to be bound, while the dibaryon with the $\Omega_{ccc}\Omega_{bbb}$ configuration is difficult to form any stable state. The study of the interaction between two full heavy baryons, and the effective potentials provide the evidence for the conclusion. Besides, the behavior of low-energy scattering phase shifts and the scattering length also confirm the existence of stable full heavy dibaryons $\Omega_{ccc}\Omega_{ccc}$ and $\Omega_{bbb}\Omega_{bbb}$. The lattice QCD studied the deuteronlike dibaryons with heavy quark flavors and with $J^{P}=1^{+}$~\cite{Lattice3}. They suggested that the dibaryons $\Omega_{c}\Omega_{cc} (sscscc)$, $\Omega_{b}\Omega_{bb} (ssbsbb)$, and $\Omega_{ccb}\Omega_{cbb} (ccbcbb)$ were stable under strong and electromagnetic interactions, and they also found that the binding of these dibaryons became stronger as they became heavier in mass. We suggest that the lattice QCD check whether the full heavy dibaryons $\Omega_{ccc}\Omega_{bbb}$, $\Omega_{ccc}\Omega_{ccc}$ and $\Omega_{bbb}\Omega_{bbb}$ with $J^{P}=0^{+}$ exist or not.

\acknowledgments{This work is supported partly by the National Science Foundation
of China under Contract Nos. 11675080, 11775118 and 11535005.}

\end{document}